\documentclass[conference]{IEEEtran}
\IEEEoverridecommandlockouts
\usepackage{cite}
\usepackage{amsmath,amssymb,amsfonts}
\usepackage{algpseudocode}
\usepackage{algorithm}
\usepackage{graphicx}
\usepackage{textcomp}
\usepackage{xcolor}
\usepackage{subcaption}
\usepackage{caption}
\usepackage{setspace}
\def\BibTeX{{\rm B\kern-.05em{\sc i\kern-.025em b}\kern-.08em
    T\kern-.1667em\lower.7ex\hbox{E}\kern-.125emX}}
\begin{document}

\title{Uncertainty Estimation in Multi-Agent Distributed Learning\\
\thanks{Funding was provided by the European Union’s Horizon 2020 research and innovation program within the framework of Key Digital Technologies  Joint Undertaking (Grant No. 101112268). }
}

\author{\IEEEauthorblockN{Gleb Radchenko}
\IEEEauthorblockA{\textit{Embedded Systems Department} \\
\textit{Silicon Austria Labs}\\
Graz, Austria \\
gleb.radchenko@silicon-austria.com}
\and
\IEEEauthorblockN{Victoria Andrea Fill}
\IEEEauthorblockA{\textit{Embedded Systems Department} \\
\textit{Silicon Austria Labs}\\
Graz, Austria \\
victoria.fill@silicon-austria.com}
}

\maketitle


\section{Introduction}

Traditionally, IoT edge devices have been perceived primarily as low-power components with limited capabilities for autonomous operations~\cite{Samie16}. Yet, with emerging advancements in embedded AI hardware design, a foundational shift paves the way for future possibilities. Thus, the aim of the KDT NEUROKIT2E\footnote{https://cordis.europa.eu/project/id/101112268} project is to establish a new open-source framework to further facilitate AI applications on edge devices by developing new methods in quantization, pruning-aware training, and sparsification. These innovations hold the potential to expand the functional range of such devices considerably, enabling them to manage complex Machine Learning (ML) tasks utilizing local resources and laying the groundwork for innovative learning approaches.

In the context of 6G's transformative potential, distributed learning among independent agents emerges as a pivotal application~\cite{Park21}, attributed to 6G networks' support for ultra-reliable low-latency communication, enhanced data rates, and advanced edge computing capabilities~\cite{Zhang19}.

Our research focuses on the mechanisms and methodologies that allow edge network-enabled agents to engage in collaborative learning in distributed environments. Particularly, one of the key issues within distributed collaborative learning is determining the degree of confidence in the learning results, considering the spatio-temporal locality of data sets perceived by independent agents.
\section{Distributed Neural Networks Training}
Distributed ML algorithms, based on their communication mechanisms, primarily exchange model parameters, model outputs, or hidden activations. These exchanges can be facilitated through peer-to-peer or client-server architectures~\cite{Park21}. The main categories into which these algorithms fall are as follows:
\begin{enumerate}
    \item \textbf{Federated Learning} --- a centralized client-server learning model, where workers periodically sync their local model's parameters via a central server. 
    \item \textbf{ADMM-derived Methods}~\cite{Boyd11} (such as GADMM, CADMM, and DiNNO~\cite{Yu22}) support decentralized learning through peer-to-peer exchange of model parameters.
    \item \textbf{Federated Distillation} --- a distributed learning method where workers share compact model outputs rather than full models, thus supporting model compression, transferring knowledge from large models to smaller ones.
    \item \textbf{Split Learning} divides multi-layer neural networks into several segments. The lower segments are placed on workers processing the raw data. During training, each worker sends activation values to the server, which calculates loss, and then return gradients for local updates.
\end{enumerate}
Bayesian neural networks (BNNs) employ a Bayesian approach to train stochastic neural networks~\cite{Jospin22}. Instead of deterministic weights and biases, they utilize probability distributions, denoted \( P(w) \) for weights and \( P(b) \) for biases. Typically, these distributions are approximated as Gaussian, with mean and standard deviation derived from the training data. Hence, a Bayesian neuron outputs a range of possible values, not just one. The operation of a Bayesian Linear neuron can be described as:
\begin{equation}
P(y|x)=f_{act}((\sum P(w))\times x+P(b))
\end{equation}
Unlike traditional NNs, which employ a singular forward pass, BNNs might conduct multiple forward passes. The mean and standard deviation of these outputs are then computed. Depending on the problem type addressed by the neural network, these mean and standard deviation values can indicate the model's uncertainty for each point in the input data space.

\section{Uncertainty Estimation for DiNNO}

In the present study, we explore methodologies to augment the DiNNO distributed learning algorithm by tailoring it for compatibility with BNNs. Our evaluation centers around a collaborative mapping case study wherein we focus on training an autonomous robot's NN with LiDAR data. Algorithm~\ref{alg:peer_sync} fosters asynchronous data interchange during the decentralized learning process among autonomous agents.

\setlength{\textfloatsep}{0pt}
\begin{algorithm}[hbt!]
\caption{Peers State Exchange}\label{alg:peer_sync}
\small
\begin{algorithmic}[1]
\Require $MaxRound, Socket, Id, State$
\State Initialize: $Round, PeerComplete[\ ], PeerState[\ ]$
\State $Message \gets (State, 0)$
\State \Call{Send}{$Socket$, $Message$, $Id$}
\While{$Round < MaxRound$}
    \State $(Message, PeerId) \gets \Call{Receive}{Socket}$
    \If{$Message$ is $RoundComplete$}
        \State $PeerComplete[PeerId] \gets \textsc{true}$
    \Else
        \If{$Round < Message.Round$} \State \Call{FinishRound}{} \EndIf
        \State $PeerState[PeerId] \gets Message.State$
    \EndIf
    \If{$\forall s \in PeerState,\ s \neq \O$}
        \State $State \gets$ \Call{NodeUpdate}{$State, PeerState$}
        \State $\forall s \in PeerState,\ s \gets \O$
        \State $PeerCompleted[Id] \gets \textsc{true}$
        \State $PeerState[Id] \gets State$
        \State $Message \gets RoundComplete$
        \State \Call{Send}{$Socket$, $Message$, $Id$}
    \EndIf
    \If{$\forall p \in PeerComplete,\ p = \textsc{true}$}
        \State \Call{FinishRound}{}
    \EndIf
\EndWhile
\Function{FinishRound}{}
    \State $\forall p \in PeerComplete,\ p \gets \textsc{false}$
    \State $Round \gets Round+1$ 
    \State $Message.State \gets State$
    \State $Message.Round \gets Round$
    \State \Call{Send}{$Socket$, $Message$, $Id$}
\EndFunction
\end{algorithmic}
\end{algorithm}

We introduce a BNN into the mapping problem, substituting the linear layers in the NN with the Bayesian Linear Layers. To ensure correct regularization of the BNN parameters, algorithm~\ref{alg:bnn_reg} is developed to take into account the semantics of median ($\mu$) and standard deviation ($\rho$) parameters of BNN neurons. We utilize Kullback–Leibler divergence for regularization of BNN $\rho$-parameters.
\begin{algorithm}[hbt!]
\caption{Optimization of BNN Parameters}\label{alg:bnn_reg}
\small
\begin{algorithmic}[1]
\Require $Model, Optimizer_{\mu}, Optimizer_{\rho}, W_{\mu}, W_{\rho}, Iter,$ 
$\theta^{\mu}_{reg}, \theta^{\rho}_{reg}, Duals_{\mu}, Duals_{\rho}$
\For{$i \gets 1$ to $Iter$}
    \State Reset gradients of $Optimizer_{\mu}$ and $Optimizer_{\rho}$
    \State $PredLoss \gets \Call{ComputeLoss}{Model}$
    \State $\theta^{\mu}, \theta^{\rho} \gets \Call{ExtractParameters}{Model}$
    \State $Reg_{\mu} \gets \Call{L2Regularization}{\theta^{\mu}, \theta^{\mu}_{reg}}$
    \State $Reg_{\rho} \gets \Call{D\_KL} {\theta^{\rho}, \theta^{\rho}_{reg}}$
    \State $Loss_{\mu} \gets PredLoss + \langle \theta^{\mu}, Duals_{\mu} \rangle + W_{\mu} \times Reg_{\mu}$
    \State $Loss_{\rho} \gets \langle \theta^{\rho}, Duals_{\rho} \rangle + W_{\rho} \times Reg_{\rho}$
    \State \Call{UpdateParameters}{$Optimizer_{\mu}, Loss_{\mu}$}
    \State \Call{UpdateParameters}{$Optimizer_{\rho}, Loss_{\rho}$}
\EndFor
\end{algorithmic}
\end{algorithm}
\begin{figure}
      \centering
	   \begin{subfigure}{0.32\linewidth}
		\includegraphics[width=\linewidth]{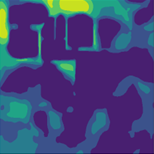}
	   \end{subfigure}
	   \begin{subfigure}{0.32\linewidth}
		\includegraphics[width=\linewidth]{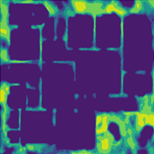}
	   \end{subfigure}
	\vfill
        \vspace{0.1cm}
	\begin{subfigure}{0.32\linewidth}
		\includegraphics[width=\linewidth]{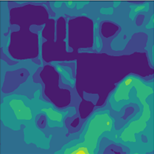}
	\end{subfigure}
	\begin{subfigure}{0.32\linewidth}
		\includegraphics[width=\linewidth]{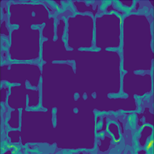}
	\end{subfigure}
	\vfill
	    \vspace{0.1cm}
	   \begin{subfigure}{0.32\linewidth}
		\includegraphics[width=\linewidth]{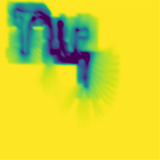}
		\caption{Single agent}
		\label{fig:subfig5}
	   \end{subfigure}
	   \begin{subfigure}{0.32\linewidth}
		\includegraphics[width=\linewidth]{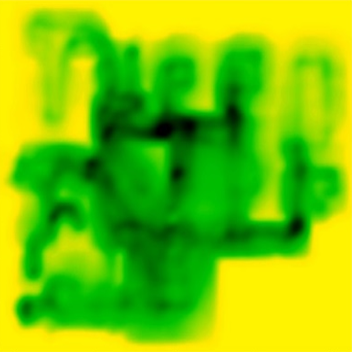}
		\caption{Multi-agent swarm}
		\label{fig:subfig6}
	    \end{subfigure}
	\caption{Comparative Visualization of Collaborative Mapping. Displaying map representation (top), uncertainty levels (middle), and LiDAR data-point densities (bottom) between single-agent~(a) and multi-agent~swarm~(b) scenarios. }
	\label{fig:subfigures4}
\end{figure}

To assess the outcomes of the BNN, we propose an uncertainty baseline using the distribution of the LiDAR data. We employ Kernel Density Estimation (KDE) to analyze the measurements and determine their inherent distribution. We utilize the multivariate kernel estimator presented in Equation~\ref{eq:kde} from~\cite{Scott15}, where $K$ is the designated kernel.
\begin{equation}
    \hat{f}(x) = \frac{1}{n|H|} \sum_{i=1}^{n}K(H^{-1}(x - x_i))\label{eq:kde}
\end{equation}
For the depiction in Figure~\ref{fig:subfigures4} the KDE is computed using the \texttt{scipy.stats.gaussian\_kde()}, which applies a Gaussian kernel. 
\section{Conclusions}
Our experiments indicate that BNNs can effectively support uncertainty estimation in distributed learning. However, one must consider the particularities of distributed learning and the required regularization to maintain high learning quality. For future work, we suggest exploring how distributed learning with BNNs can be tailored for embedded AI hardware. This would involve refining the NN architecture to suit the resource constraints of edge devices. Additionally, understanding and optimizing the network load for edge devices is crucial, given the potential of upcoming network infrastructures like~6G.

\end{document}